\documentclass[sigconf]{acmart}

\usepackage{booktabs}
\usepackage{graphicx}
\usepackage{caption}
\usepackage{subcaption}
\usepackage{url}
\usepackage{multirow}
\usepackage{algorithm}
\usepackage{algpseudocode}
\usepackage{mathpazo}
\graphicspath{ {figs/} }
\usepackage{textcomp}

\setcopyright{rightsretained}

\copyrightyear{2018} 
\acmYear{2018} 
\setcopyright{acmcopyright}
\acmConference[SAC 2018]{SAC 2018: Symposium on Applied Computing }{April 9--13, 2018}{Pau, France}
\acmPrice{15.00}
\acmDOI{10.1145/3167132.3167318}
\acmISBN{978-1-4503-5191-1/18/04}

\begin{document}
\title[Detecting Twitter Users with Manipulated Follower Count]{The Follower Count Fallacy: Detecting Twitter Users with Manipulated Follower Count}

\author[A. Aggarwal]{Anupama Aggarwal}
\affiliation{%
  \institution{IIIT - Delhi}}
\email{anupamaa@iiitd.ac.in}

\author[S. Kumar]{Saravana Kumar}
\affiliation{%
  \institution{CEG, Guindy}}
\email{shanmugamsakthivadivel.1@osu.edu}

\author[K. Bhargava]{Kushagra Bhargava}
\affiliation{%
  \institution{IIIT - Delhi}}
\email{kushagrab@iiitd.ac.in}

\author[P. Kumaraguru]{Ponnurangam Kumaraguru}
\affiliation{%
  \institution{IIIT - Delhi}}
\email{pk@iiitd.ac.in}


\begin{abstract}
Online Social Networks (OSN) are increasingly being used as platform for an effective communication, to engage with other users, and to create a social worth via number of likes, followers and shares. Such metrics and crowd-sourced ratings give the OSN user a sense of social reputation which she tries to maintain and boost to be more influential. Users artificially bolster their social reputation via black-market web services. In this work, we identify users which manipulate their projected follower count using an unsupervised local neighborhood detection method. We identify a neighborhood of the user based on a robust set of features which reflect user similarity in terms of the expected follower count. We show that follower count estimation using our method has 84.2\% accuracy with a low error rate. In addition, we estimate the follower count of the user under suspicion by finding its neighborhood drawn from a large random sample of Twitter. We show that our method is highly tolerant to synthetic manipulation of followers. Using the deviation of predicted follower count from the displayed count, we are also able to detect customers with a high precision of 98.62\%. 
\end{abstract}

\ccsdesc[500]{Information systems~Web searching and information discovery}
\ccsdesc[500]{Information systems~Social networks}

\keywords{Online Social Networks, User Behavior, Social Reputation}

%
%
%
%
%

\maketitle

\section{Introduction} \label{sec:introduction}
	With the growing rise of Online Social Networks (OSNs), services like Facebook, Twitter, and Instagram, have emerged as platform for mass communication, used by various brands, celebrities, and political parties to engage with users. The follower count and crowdsourced ratings give the OSN user a sense of social reputation which she tries to maintain and boost to be more influential in the network and attract more following. Various OSNs have a different measure of user following and social reputation, like \texttt{followers} on Twitter and Instagram, \texttt{likes} on Facebook and \texttt{ratings} on Yelp. However, these reputation metrics can be manipulated in several ways. 

	One of the most prevalent methods to alter social reputation is online black-market services, which help the users to increase their follower/like count. There exist several online services from where an OSN user can purchase bulk \texttt{followers} and \texttt{likes}. The follower black-market services also often have a collusion network model where a user can gain followers for free by following other customers of the service. Users exploit these services to inflate their social media metrics such as -- followers, likes and shares (of the user post) in the hope to become more influential and popular on the network. 

	There are several adverse effects of social reputation manipulation. Crowdsourced manipulation is a big threat to the credibility of e-commerce networks that rely on user ratings and reviews for product recommendation and build user trust. Recently Amazon sued 1,114 sellers on Fiverr -- a micro-task supply driven marketplace, for posting fake reviews of Amazon products~\cite{NYTimes:2015dq}. Such marketplaces may also create fake OSN accounts to be served as followers and likers for their customers. Infiltration of fake accounts and metrics also has a damaging effect on the advertisement revenue framework of OSNs. An advertisement might be shown to certain users with seemingly high popularity and reputation, but will most likely not get clicked by the expected number of real users due to the accounts being fake or bots~\cite{de2014paying}. Social reputation manipulation also affects the in-built recommendation services of Online Social Networks which are used to enhance user experience, like -- Who-to-Follow recommendation, top tweets during an event on Twitter, suggested pages and posts on Facebook and highest pinned boards on Pinterest. Previous studies have shown that an untrustworthy user (with bot activity) can have high influence if it has high popularity (like follower count on Twitter)~\cite{aiello2012people}. Therefore, follower count manipulation can have a damaging effect on some popular social influence metrics like Klout~\footnote{\url{http://klout.com}} and Twitlyzer~\footnote{\url{http://www.twitlyzer.com}}. These tools are vulnerable to simple automatic strategies and can be easily affected by the presence of fake metrics~\cite{messias2013you}. We found that 78.3\% customers (users with manipulated follower count) of Twitter follower black-market service in our dataset have a Klout score greater than 40 (average Klout score). This shows the implication of manipulation on influence metrics, and highlight the need of detecting such manipulation.
	
	
	In this work we study the behavior of customers (Twitter users) of black-market services that obtain inorganic followers. While most of the previous studies attempt at finding the inorganic following behavior and Sybil users responsible in increase of follower count, we focus on identifying the users with manipulated follower count and also project an estimate of real follower count which is tolerant to inorganic gain of followers. We build a framework based on following intuitions--
	
	(1) \textbf{\emph{Temporal signatures are hard to manipulate.}} While users with manipulated follower count can resemble a legitimate profile very closely, the gain in followers for such users is often in bursts. Similarly, users with bot tweeting activity (due to control by black-market they are subscribed to) might maintain their tweets/follower ratio to evade automatic detection mechanism. However, the bot tweeting activity can be uncovered by checking the periodicity of time of posts. Therefore, temporal signatures can be a strong indicator of inorganic follower gain. 
	(2) \textbf{\emph{Follower count of a user should be similar to its neighbors.}} We use several features to identify similar users from the local neighborhood of the user, sampled from a large set of random Twitter users. We define neighborhood as set of users with similar social standing, temporal trace and social signature. The details of neighborhood detection of a user is explained in Section~\ref{sec:prediction}. We then check the deviation of follower count of the user from its neighbors to identify whether it is a customer (with manipulated follower count) or not. We also project a more realistic follower count for the user based on its neighborhood which is tolerant to manipulation and give details of our results in Section~\ref{sec:estimation}. In this work, we present the following contributions
	\begin{itemize}
	\item We extract temporal signatures to identify customers of black-market services. 
	We show that features reflecting the temporal evolution of a user profile are tolerant to manipulation, and hence good indicators to find users with manipulated follower count.
	\item We devise an unsupervised nearest neighbor detection for Twitter users to discover other similar users in terms of their social standing. We leverage the neighbors of a user to find whether the user has a manipulated follower count or not.
	\item While most of the studies stop at detecting users with manipulated follower count, we also estimate the untampered follower count of a user based on its neighbors. This enables us to project a follower count of the user which is closer to the untampered value and is also tolerant to manipulation.   
	\end{itemize}

\section{Background and Challenges} \label{sec:background}
\paragraph{\textbf{Background:}} There exist several services to gain Twitter followers which users can subscribe to. These services are either paid or free. In the paid model, customers can purchase followers by making a payment to the marketplace. Some of these marketplaces offer 1000 followers for as cheap as 3 USD. On the other hand, there also exist a freemium model offered by these marketplaces which thrives on a collusion network of followers. The customer needs to give away their Twitter credentials, or use the third-party app of the marketplace to subscribe to the freemium services. In this model, customer gains followers by following other users. 
In such cases, customers become part of the collusion network of followers and are sometimes also used to advertise about the subscribed black-market service. 

\vspace{-0.25cm}
\paragraph{\textbf{Challenges:}} The customers subscribed to freemium marketplaces and part of collusion network can be hard to identify because -- (1) Due to controlled activity, user properties like follower/friends ratio and tweets/follower ratio of customers is manipulated by marketplaces to evade detection, and is very similar to a random Twitter user, (2) The followers gained by these customers are often other customers subscribed to the same service. Hence these followers are real Twitter users as opposed to bulk accounts created to cater to customers of paid services. Previous work based on the assumption of bulk account creation to detect anomalous follower gain by looking at the creation time of followers is able to detect only 22.3\% customers of freemium marketplaces in our dataset~\cite{viswanath2015strength}, (3) Although previous studies have labeled freemium customers as \emph{victims} and identified them by looking for promotional content in their tweets~\cite{stringhini2013follow}; we find less than 30\% of freemium market customers posting advertising content. This could be due to manual deletion of tweets, or because not all black-market services use their customers for promotion. Hence, presence of promotional content can not be a reliable indicator to detect these customers. 

\section{Related Work} \label{sec:related}
\paragraph{\textbf{Detecting Sybil and fake accounts used for manipulation.}} Recent studies have shown that the black-market services used to gain follower/likes create fake identities in bulk to cater to the customers~\cite{motoyama2011dirty, wang2012serf}. These fake identities infiltrate the social network for various purposes like, manipulating Twitter follower count~\cite{cresci2015fame, thomas2013trafficking}, e-commerce product reviews~\cite{luca2016fake} or Facebook likes~\cite{viswanath2014towards}. There have been several efforts made to identify the entities which cause crowdsourced manipulation. Primarily, research has focused on identifying difference in characteristics of Sybil and benign entities~\cite{benevenuto2010detecting, lim2010detecting, wang2013you, wang2014man}. Studies have also used unsupervised learning~\cite{viswanath2014towards, kumar2017fairjudge} and trust-propagation techniques to identify Sybils and fake accounts~\cite{yu2006sybilguard, yu2008sybillimit, savage2015detection}. In this work, instead of detecting specific Sybil entities, we focus on detecting users which have manipulated follower count.

\paragraph{\textbf{Detecting crowdsourced manipulation.}}
While there have been several efforts done towards finding Sybils and fake entities, there exist few studies which detect crowdsourced manipulation, like fake follower count, likes or products with synthetic reviews. Our focus in this work is to detect such crowdsourced manipulation, more specifically - Twitter follower count. Crowdsourced manipulation of discussions in online forums has been studied in past~\cite{kumar2017army}; and previous work has explored detection of manipulated e-commerce product ratings by examining the ratings of a product with other products whose ratings are untampered~\cite{feng2012distributional, wu2010distortion}, and uncover crowdsourced manipulation in large-scale by the use of temporal traces~\cite{viswanath2015strength, cao2014uncovering, beutel2013copycatch}. Previous studies based on temporal traces assume that the crowdsourced manipulation happens in bulk and by accounts created around the same time~\cite{viswanath2015strength}. We do not make any such assumption, because the customers we are studying belong to a collusion network. Hence, the gained followers might not be fake users created in bulk by the black-market, but real users which are part of the collusion network. We explore other temporal patterns and signals using which we can identify hard-to-detect customers without making any assumptions about bulk creation of fake followers.  

\paragraph{\textbf{Defence against manipulation}} 
The current defense mechanism against Sybils and crowdsourced manipulation involves suspension of accounts~\cite{thomas2011suspended, parkinson2014instagram}. There have been very less efforts made to estimate the real worth of a user which might have manipulated follower/like count. There have been attempts in past to measure influence of an OSN user, like using an influence metric called Klout~\cite{rao2015klout}. While Klout gives high ranking to an influential and a highly recognizable user, it is not designed to be tolerant to crowdsourced manipulation of metrics such as followers and likes. In addition, it does not give a sense of real follower count of a user. In this work, we take a step forward in this direction, and predict the follower count of users using metrics which are tolerant to manipulation. 

\section{Data Overview} \label{sec:data}
In this work, we study users which manipulate their Twitter follower count by subscribing to free services provided by black-markets. There exist several Twitter follower marketplaces. To collect ground-truth data of users which manipulate their Twitter follower count, we extract customers of these markets. We picked top 6 websites based on their availability of service and their global ranking on Alexa~\footnote{Alexa. \url{http://www.alexa.com/}}. 
In total we collect 38,791 customer users and call this dataset of customers D$_{cust}$. We notice that a substantial fraction (16\%) of the customers were subscribed to more than one service. In further analysis of customers, we compare their behavior with that of random users on Twitter in Section~\ref{sec:comparison}, for which we also collect 127,922 random Twitter users and call this dataset D$_{rand}$. 


To study temporal behavior of users, we collect user information of customers and random users after every 6 hours (total 30 snapshots, spanning across a week~\footnote{We restrict to one week due to Twitter API rate limit.}) and call these datasets D$_{cust}^{temp}$ and D$_{rand}^{temp}$ respectively. Table~\ref{tab:dataset-details} gives details of the information we collect for the customers and random users, which is -- (1) temporal profile information of users, (2) static list of followers and their profile information, (3) temporal list of friends, (4) most recent 200 tweets (maximum number of tweets returned in a single API call) of all customers and random users. 

\begin{table}[h]
\centering
\small
\begin{tabular}{@{}lllll@{}}
\toprule
            & \multicolumn{2}{l}{Customer} & \multicolumn{2}{l}{Random } \\ \midrule
            & D$_{cust}^{temp}$       & D$_{cust}$       & D$_{rand}^{temp}$      & D$_{rand}$      \\ \midrule
\#Users     & 26,917              & 38,791           & 14,788             & 127,922          \\
\#Followers & --              & 24,514,951           &  --             & 10,619,990          \\
\#Friends   & 12,064,797              &  --           & 43,335,556             &  --          \\
\#Tweets    & --              & 4,134,854           & --             & 15,016,287          \\ \bottomrule
\end{tabular}
\caption{We collect temporal information of a sub-sample of customers and random users. D$_{cust}$ and D$_{rand}$ represent the static data, and D$_{cust}^{temp}$ and D$_{rand}^{temp}$ represent the temporal trace of customers and random users.}
\label{tab:dataset-details}
\end{table}

\begin{figure}[h]
\centering
\begin{subfigure}{.23\textwidth}
  \centering
  \includegraphics[width=.9\textwidth]{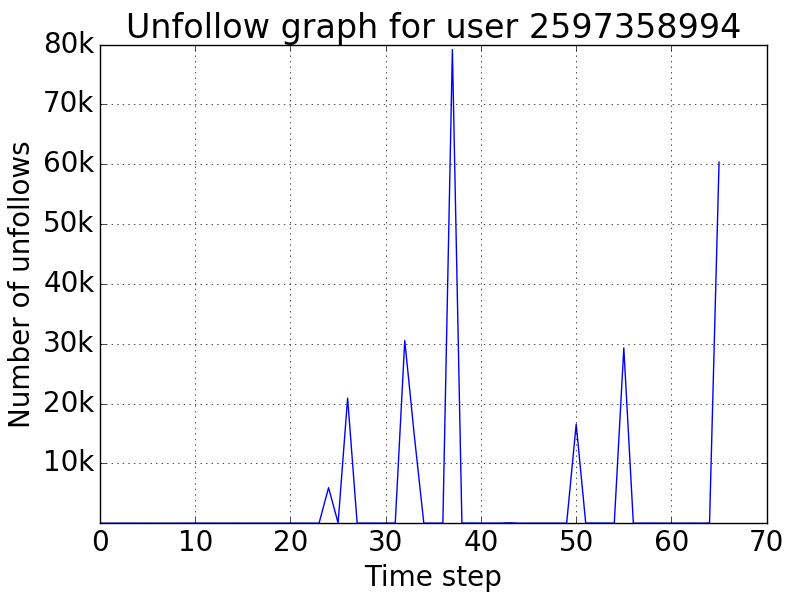}
  \caption{Sample user with multiple unfollow instances}
  \label{fig:unfollow_2597358994}
\end{subfigure}%
\begin{subfigure}{.24\textwidth}
  \centering
  \includegraphics[width=.9\textwidth]{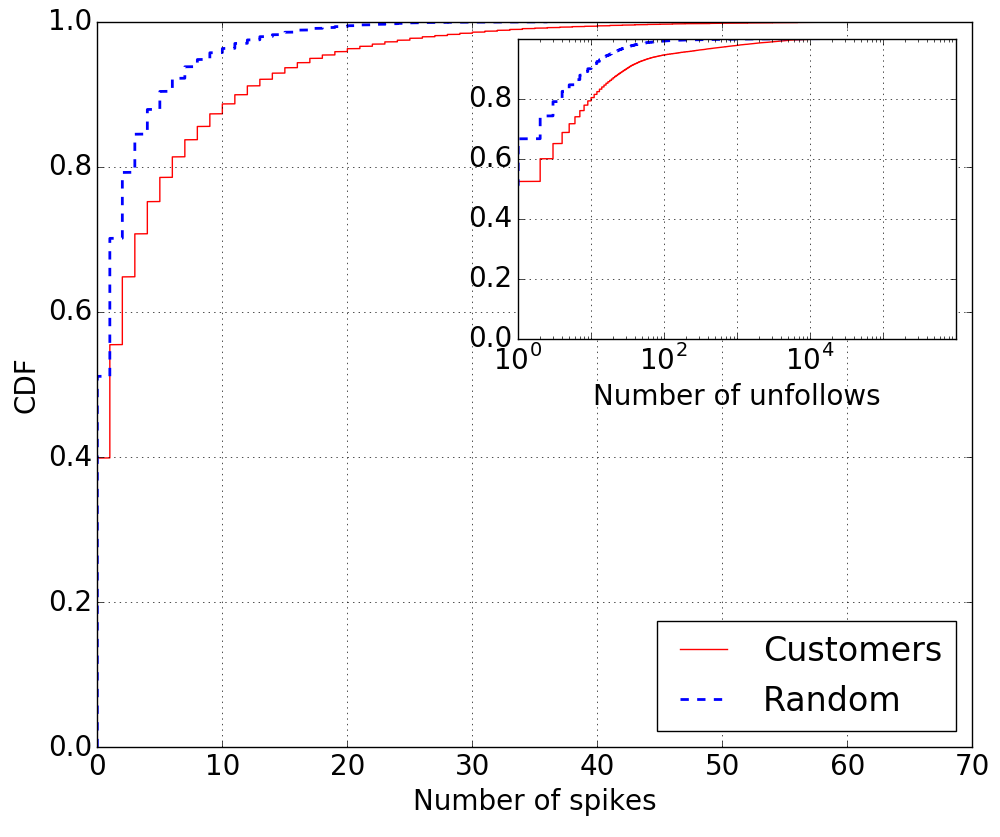}
  \caption{Number of unfollow instances by customers}
  \label{fig:num_unfollow_spikes}
\end{subfigure}
\caption{Customers with aggressive unfollow activity}
\label{fig:test}
\end{figure}

Since the customers of the black-market services which we collect are subscribed using the freemium model, they are part of the collusion network where they follow other users (often controlled by black-market service) in return of followers. We notice that the customers often unfollow a fraction of their friends and follow them back, and repeat this activity several times with a non-unique set of their friends. We call such behavior \emph{unfollow activity}. To identify customers which use black-market services aggressively, we closely monitor the frequency of unfollow activity by these users. Some users unfollow their friends as much as 27,038 times with a repeated unfollow activity within a span of a week as shown in Figure~\ref{fig:unfollow_2597358994}. We give details of such behavior across the customer dataset in Section~\ref{sec:comparison}. Figure~\ref{fig:num_unfollow_spikes} shows the CDF of number of spikes in unfollow count over time. We use this to identify a subset consisting of 12,565 customers which exhibit aggressive unfollow activity, that is, unfollow their friends more than once, and call this dataset D$_{agg}$. 

\section{Customers vs. Random Users} \label{sec:comparison}
Our goal is to detect follower count manipulation on Twitter. To achieve this, we first extract features based on temporal and social signatures which can help identify manipulation. We perform a thorough comparison of users profiles of customers of black-market services and a set of random Twitter users.

\begin{figure}
\centering
\includegraphics[width=0.48\textwidth]{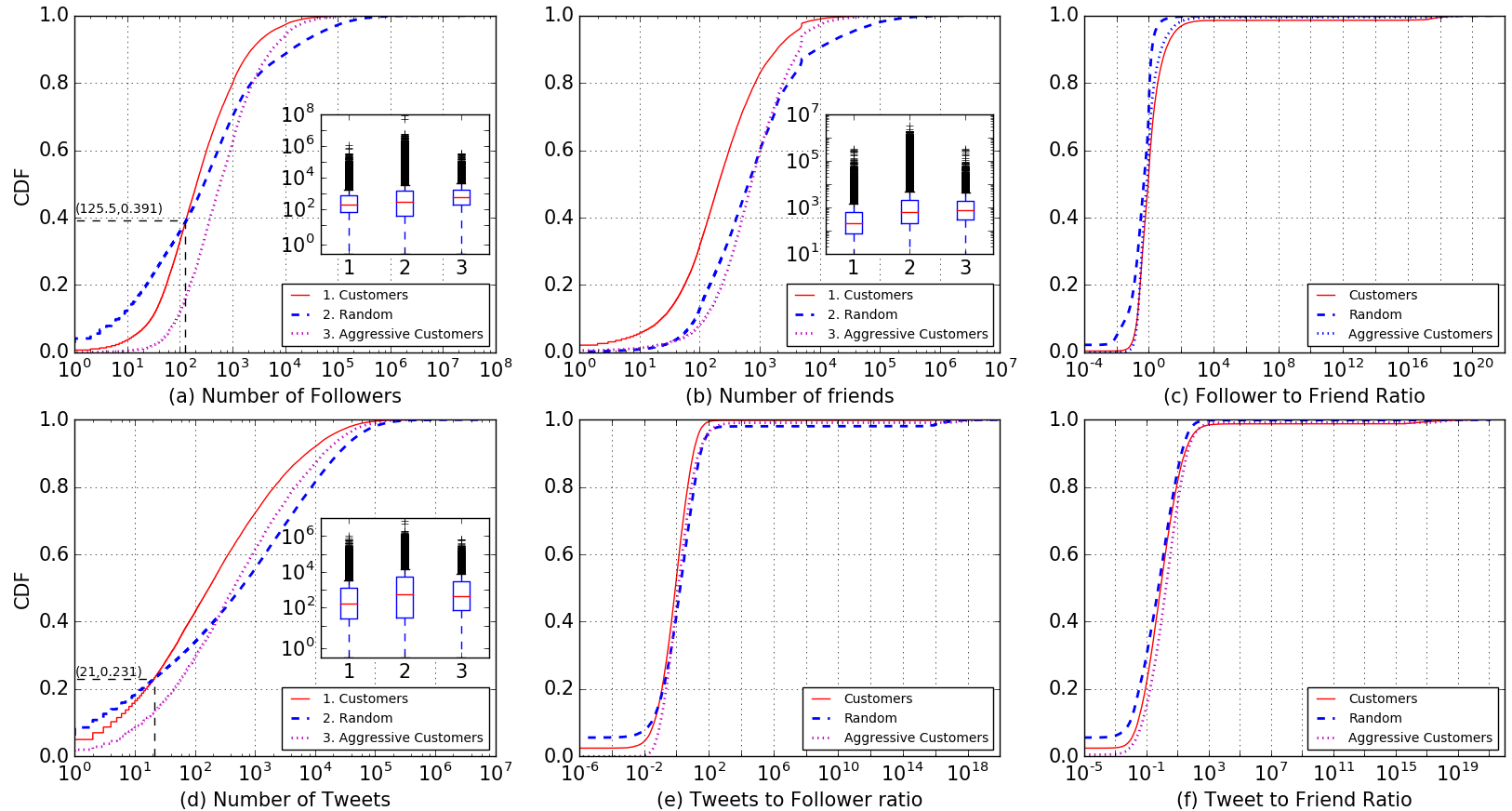}
\caption{Static features comparing customers with random Twitter users. The CDFs show close similarity of customers and random users with respect to the above features.}
\label{fig:staticfeatures}
\end{figure}

\subsection{Static Feature Analysis} \label{subsec:static}
We use D$_{cust}$ and D$_{rand}$ datasets consisting of 38,791 and 127,922 number of users in customer and random datasets. We capture the profile information of customers and random users at the last snapshot of the temporal data collected. We do not guarantee that a fraction of users in D$_{rand}$ would not have manipulated their follower count. However, since we take a random sample of users from Twitter which is substantially larger than the customer dataset, we argue that these two set of users are inherently different. 

As seen in Figure~\ref{fig:staticfeatures}, the CDF distributions of static features of customers and random users based on follower count, friends count and tweet count have non-significant difference. Such proximity in the profile based features in case of black-market customers and random users shows that the count of followers, friends and tweets can be easily manipulated by customers, or controlled by marketplaces to evade detection.  

The CDF of follower count in Figure~\ref{fig:staticfeatures}a shows that till 125 number of followers, a smaller fraction of random users have as much followers as customers. However, after crossing the threshold of 125, a larger fraction of random users have higher follower count as compared to customers. Since customers of freemium marketplaces gain followers in small bursts, 125 is close to the average number of followers gained by customers per day. This explains why a smaller fraction of random users have as much followers as customers till this threshold. However, since the gain in followers in case of freemium customers is slow, the genuine celebrity users in random dataset of Twitter users have a higher follower count above the 125 threshold of followers; which explains the upper portion of the CDF. Similarly, the friends count distributions of customers and random users in Figure~\ref{fig:staticfeatures}b are closely associated. 

We notice, that in Figure~\ref{fig:staticfeatures}c that the friends count and follower/friends ratio of the two distributions is almost overlapping. We further analyse the difference in the number of tweets posted and tweet/follower ratio to assess if the the frequency of content can be used as proxy to user popularity and follower count. However, we find that tweet count, and CDF of tweet/followers and tweet/friends of customers and random users is highly overlapping (Figure~\ref{fig:staticfeatures}d--~\ref{fig:staticfeatures}f). We also analyse the other static features of the users at a single time snapshot and find out that 1.72\% customers have egg profiles (profiles with no display image) as compared to 9.2\% egg profiles in random user dataset. Also, we did not find any significant differences in bio of the two datasets, with an average bio length of 56.19 characters for customers and 51.39 characters for random users. 

This makes the detection of customers a challenging task and eliminates the possibility of effective detection based only on the static features. Therefore, we move towards finding temporal signatures of users based on their tweeting activity, profile evolution over time, and the trace of their following behavior in the forthcoming analysis. 


\subsection{Posting Patterns} \label{subsec:posting}
We extract the first set of temporal trace for users from their most recent 200 tweets. Our intuition behind finding temporal signatures from tweets is that though marketplaces can control the number of tweets and maintain the ratio with friends and followers, the time of tweets can reveal anomalous bursts in tweeting activity after the customer subscribes to the black-market service. 

\paragraph{\textbf{Analysing Post Content}} We assess the content posted by the customers in comparison to random users. We find that the fraction of tweets which are retweets of other users is similar in case of customers and random users with 65\% customers having half of their content as retweet on an average vs 72\% of random users. We also analyse if customers engage more or less with other users, by looking at the fraction of tweets which are replies or mention other users (by searching for @mention tag). Though one may expect customers to be less engaging with other users, we surprisingly find out that on an average, customers engage as much as any random Twitter user. We also find customers using similar number of hashtags and URLs in their tweets as a random Twitter user. We believe that such behavior is due to the fact that the customers of freemium market which we are studying do not exhibit large-scale bot activity in their tweeting behavior. Instead, these are real users which engage in manipulation of their follower count by leveraging collusion network. Therefore we can not find explicit difference in their posting pattern (as used in previous studies for bot detection~\cite{zhang2011detecting}), which proves to be another challenge to detect such customers. 

\begin{figure}[h]
\centering
\includegraphics[width=0.45\textwidth]{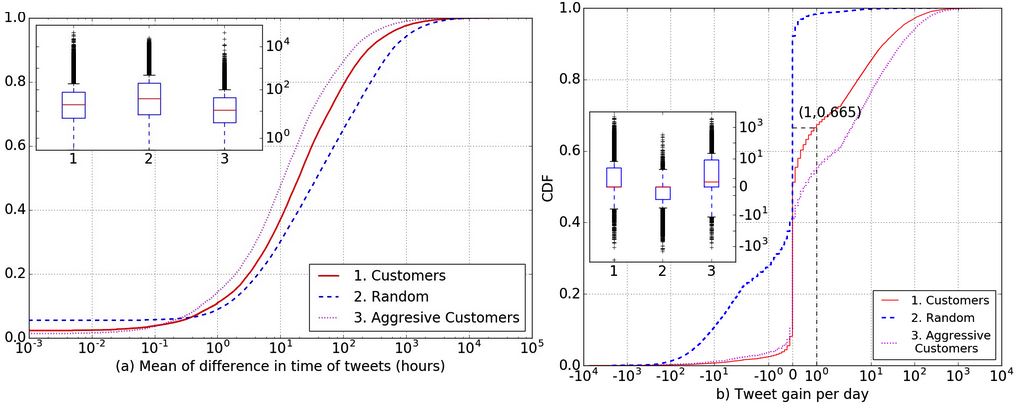}
\caption{Analysing the time of tweets and tweet gain per day of customers vs random users}
\label{fig:postingtimemean}
\end{figure}

\paragraph{\textbf{Analysing Time of Post}} To compare how frequently customers post as compared to random users, we look at the mean difference between two consecutive tweets by looking at past 200 most recent tweets of all users. We find that there is only a non significant difference in the average time between two consecutive posts by customers and random users (Figure~\ref{fig:postingtimemean}a). In addition, we also analyse how many tweets are posted by the user per day. We find out that in case of random users and customers, the average gain in tweets is similar. However, we find difference in the temporal patterns of time of posts. 33.5\% of customers post at least one tweet per day as compared to only 0.25\% of random users as shown in Figure~\ref{fig:postingtimemean}b. Though the average difference in posts of customers is similar to random users, since the customers might be posting multiple tweets in bursts, the tweet gain per day of the customers is significantly higher than that of random users. However, we also find a small fraction of customers deleting tweets. On further investigation, we found that deleted tweets were the promotional tweets by black-market services. This analysis shows that carefully crafted temporal signatures can be difficult to evade detection and hence can be distinguishing factor to identify customers of collusion network.

In addition, to assess if the customers exhibit any kind of bot tweeting activity, we look for periodicity of time of tweets to find if there are specific time snapshots when the customers post tweets. We do not find any significant traces of bot tweeting activity. 


	
\subsection{Longitudinal Temporal Analysis} \label{subsec:temporal}
To successfully find difference in behavior of customers we first collect a week long temporal trace of following activity of customers and a subset of random users by taking multiple snapshots of their friends and gain in followers each day. We use the previously described datasets of D$_{cust}^{temp}$ and D$_{rand}^{temp}$. In addition, we take a subset of customers, which is D$_{agg}$, comprising of customers exhibiting aggressive (and repeated) unfollow activity towards their friends. For this part of study, we focus on the following aspects -- (1) follower gain over time, (2) \emph{unfollow activity} -- following and unfollowing the same user repeatedly, (3) friend gain over time.

\begin{figure}[h]
\centering
\includegraphics[width=0.4\textwidth]{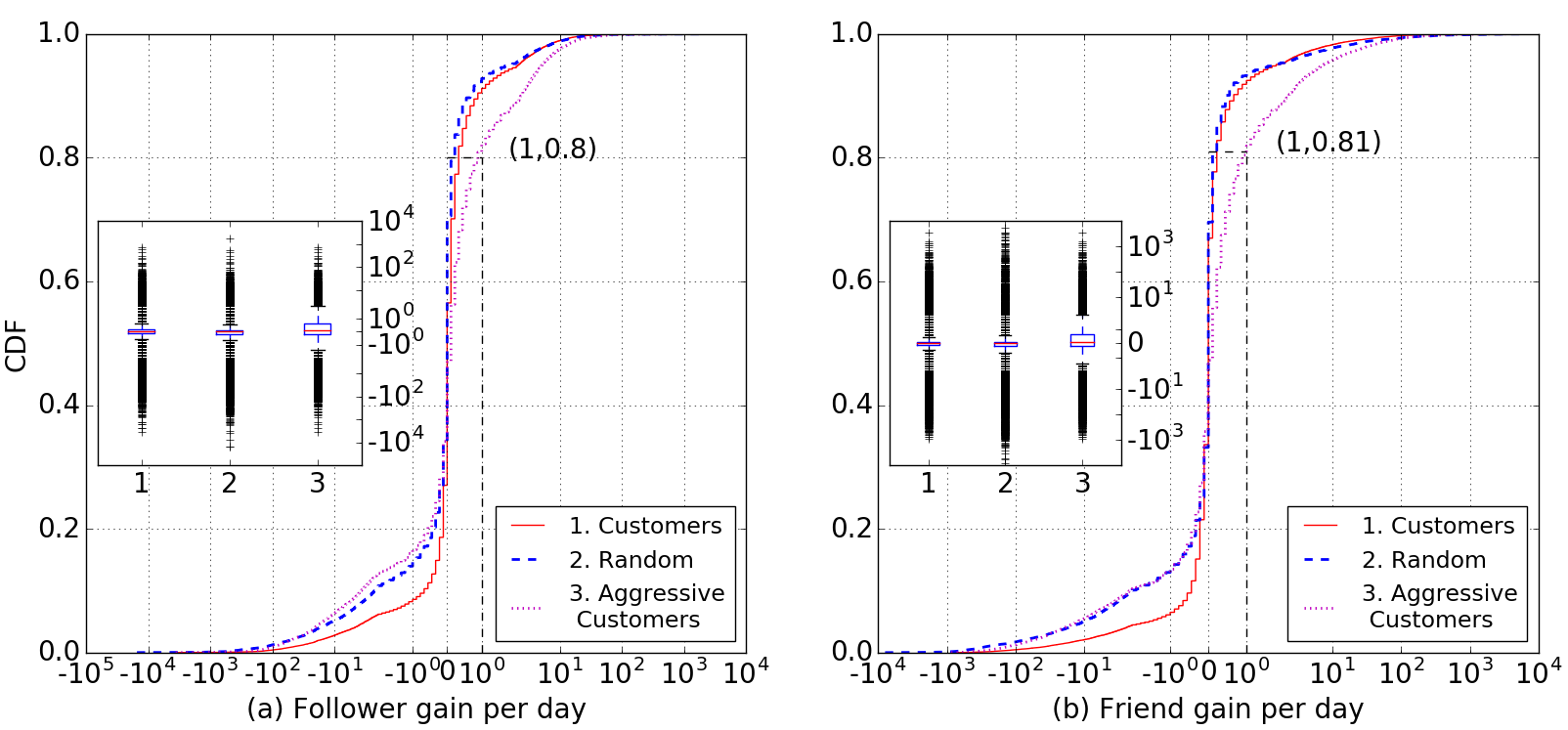}
\caption{Temporal gain in followers and friends for customers and random users}
\label{fig:followerfriendgain}
\end{figure}

\paragraph{\textbf{Temporal Follower Gain: }} Since the customers manipulate their follower count by using black-market services, we expect them to have larger follower gain per day as compared to random users. Figure~\ref{fig:followerfriendgain}a shows that there is a significant difference between the follower gain per day in case of aggressive customers and random users. 80\% aggressive customers have an average of one follower gain per day (or less) as compared to only 0.05 (or less) followers gained per day by same fraction of random Twitter users. This clearly shows the abnormal growth in follower count in case of customers. We also observe that a 49\% customers have a significantly lesser cumulative gain in followers as compared to random users. This could be because they do not take regular services from the black-market websites and indulge in follower count manipulation at a small scale only few number of times. However, for aggressive customers, the follower gain per day is significantly different from random users as shown in Figure~\ref{fig:followerfriendgain}a. We also notice that about 16.8\% customers lose followers. These could be customers which unsubscribe from black-market services and hence start losing followers. A deeper analysis of these customers showed that they gained less than 50 followers in span of a week. Since follower gain is a distinguishing factor, we use average follower gain as one of the metrics to determine abnormal growth, and to also find out similar users with respect to their unmanipulated follower count.

\begin{figure}[h]
\centering
\includegraphics[width=0.45\textwidth]{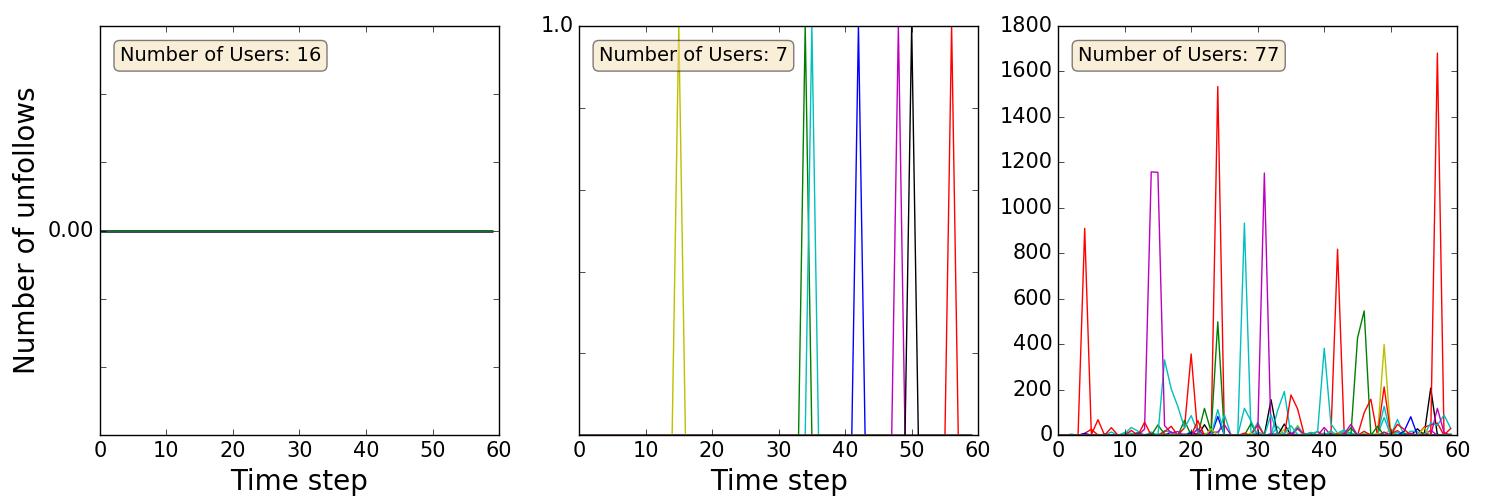}
\caption{Time series clustering of users with repeated unfollowing patterns}
\label{fig:timeseriescluster}
\end{figure}

\paragraph{\textbf{Unfollow Activity: }} We now study the aggressive customers in more detail. Aggressive customers are a subset of customers which repeatedly follow a set of friends which they have previously unfollowed. To determine aggressive customers, we cluster the customers based on their time time series data of how many previously following friends do customers unfollow. Figure~\ref{fig:timeseriescluster} shows the clustering result of a sample of 100 randomly picked customers. Based on spectral time clustering, we get three clusters. The first cluster consists of customers which do not exhibit any unfollow activity. The second cluster contains the customers which follow back previously unfollowed friends only once. And the third cluster in the figure shows customers which repeatedly unfollow their friends and follow them back, which we call aggressive customers (since they unfollow their friends several times). We repeat the spectral time clustering over the entire dataset of customers and find 12,565 users in D$_{agg}$, that is, the aggressive customer dataset. To capture the repeated unfollow activity, we use previously defined unfollow entropy of a Twitter user~\cite{aggarwal2015they}, where a larger entropy indicates higher amount of repeated unfollowing activity. 
We find that 97.2\% customers have a very high entropy of 0.78, as compared to the average unfollow entropy of only 0.12 for the random Twitter users. 

\paragraph{\textbf{Temporal Friend Gain: }} Figure~\ref{fig:followerfriendgain}b shows that aggressive customers have a significantly more friend gain that random users, with 81\% customers following at least one new user every day. This shows that despite unfollowing a large number of friends, customers gain friends over time at a higher rate than other Twitter users. One of the possible reasons for such behavior could be the collusion network of which the customers in our study are part of. In order to steadily gain followers, the customers have to follow other users (manually, or controlled by marketplaces). One of the possibility towards such behavior is that the marketplace's third party app makes the customer unfollow a set of users repeatedly to maintain the overall friend/follower ration to evade detection. The second plausible cause of such behavior is that the customers manually keep unfollowing the users which they are made to follow by the marketplace's third party app once they notice that their account is being used to follow other users, whose content they are not interested in. This is likely the reason for 17.7\% customers to have a negative follower gain (as compared to 22\% random users). 


\vspace{-0.15cm}
\section{Predicting Follower Count} \label{sec:prediction}

Our follower count prediction methodology is based on the intuition that users of similar social standing and behavior will have similar follower count. Now to determine user similarity which can reflect similar follower count, we define a user using a multi-dimensional vector of a set of features based on the deterministic signals we elaborated in the previous section. Based on these features, we find proximity of the user under concern with a large sample of random Twitter users to find its local neighborhood of users. We predict the follower count of the user by taking mean follower count of the users in it's neighborhood. 

\subsection{Features for User Similarity}
We now elaborate the features based on which we define a user as multi-dimensional vector. We divide our features into three parts based on previous analysis, viz. static features, posting behavior and temporal trace. Table~\ref{tab:features} lists the details of features which we use for defining a user. In addition to the previously described features, we also add the following --

\emph{Celebrity user: } We determine whether a user is celebrity or not by first checking if it is verified by Twitter. Next, we look up Google Knowledge Graph API \footnote{\url{https://developers.google.com/knowledge-graph/}} to determine if the user has a known Wikipedia page. Either if the user is verified, or has a Wikipedia page, we consider the user to be a celebrity. This feature helps us to measure the follower count of a user by taking into account the popularity of that user.

\emph{Topics of interest: } We extract topics a user is interested in from the bio of the user. The list of extracted topics can be effective indicator of similar users with the one for which we want to predict the follower count. We also determine the overlap of topics extracted from bio of the user with other users in the random Twitter dataset to find similar users.

\emph{Spikes in creation time and follow time: } Customers of freemium marketplace which we are studying gain followers in short bursts. These bursts can be captured by determining the spikes in number of followers gained per day. In addition, a fraction of the gained followers can be fake and stockpiled accounts controlled by the black-market operator. Hence, the creation time of such fake accounts would be similar which we capture by determining the spikes in the number of followers created on the same date. Previous studies have shown that such method can effectively find follower count manipulation in certain cases~\cite{viswanath2015strength}.  

\begin{table}
\centering
\small
\begin{tabular}{lp{5.5cm}}
\toprule
Static Features   & Presence of bio, presence of profile pic, number of tweets, friends count, listed count, is verified or not, is celebrity or not, topics of interest, topic overlap with followers, favorited tweet count \\ \midrule
Posting Behavior  & Tweet gain, tweet language, language overlap with followers                                                                                                                                               \\ \midrule
Temporal Features & Follower gain, friends gain, spikes in creation time of followers, spikes in follow time of followers, unfollow entropy \\ \bottomrule                                                                                 
\end{tabular}
\caption{Features used for determining follower count similarity \& define user as a multi-dimensional feature vector}
\label{tab:features}
\end{table}

\vspace{-0.3cm}
\subsection{Follower Count Prediction based on Local Neighborhood Detection}
Algorithm~\ref{nnfollower} describes the method we use to predict follower count of a user without taking into account the displayed follower count on its profile which could be manipulated. To determine the follower count of a user, we first choose a local neighborhood of the user from a large sample of random Twitter users by determining N-nearest neighbors of the user. Each user is defined by the feature vector described in the previous section in Table~\ref{tab:features}. The N-nearest neighbors are determined by finding the distance between the user vector for which we want to predict the follower count, and the user vectors belonging to the random Twitter dataset. We experiment with KDTre and BallTree nearest neighbor algorithms. We determine the number of relevant neighbors for a user by taking into account only the users with whom proximity of the user falls within the first and the last quartile value. Using inter-quartile range to avoid outliers, is a standard method in statistics~\cite{walfish2006review}. This ensures that we do not count any outliers as relevant neighbors for the user. Next, based on these users we determine the follower count of the user under observation by taking weighted mean follower count of the neighbors, where the weight is indirectly proportional to the distance of neighbor from the user. Weighted mean reduces the effect of less similar users and the most proximal users are given a higher priority to contribute towards the predicted follower count. Since the neighbors are from a large representative set of random Twitter users, we argue that the real follower count of the user should be close to the other users in its proximity.

\begin{algorithm}
\small
\caption{Follower Count (f) Prediction of Twitter user (u)}\label{nnfollower}
\begin{algorithmic}[1]
\Procedure{FollowerCount}{$u,f$}
\State $\vec{u}\gets$ Feature vector as described in Table~\ref{tab:features} for user $u$
\State R = $\{r_1, r_2...r_n\}$ \Comment{Random Twitter users} 
\State DistR = \{\} \Comment{Dict. containing distance of $r_i$ from u}
\State N = \{\} 	\Comment{Dict. containing neighbors of u}
\ForAll{$r_i$ in R}
	 \State $\vec{r_i}\gets$ Feature vector for user $r_i$
	 \State d($u, r_i$) $\gets$ $|\vec{u} - \vec{r_i}|$ \Comment{Distance of u from $r_i$} 
	 \State DistR[$r_i$] $\gets$ d($u, r_i$) 
\EndFor
\State RankedDistR $\gets$ sort(DistR) \Comment{Sorted by distance} 
\State X$_l$ $\gets$ LowerQuartile(RankedDistR)
\State X$_u$ $\gets$ UpperQuartile(RankedDistR)
\State IQR $\gets$ X$_u$ - X$_l$
\ForAll{$r_i$ in RankedDistR}
	\If{\small{X$_l$ - 1.5*IQR $\leq$ RankedDistR[$r_i$] $\leq$ X$_u$ + 1.5*IQR}} 
		\State N[$r_i$] $\gets$ RankedDistR[$r_i$] 
	 \EndIf
\EndFor
\State f = $\frac{1}{\sum\limits_{j=1}^{|N|} d_j}$$\sum\limits_{i=1}^{|N|} \frac{f_i}{d_i} $ \Comment{Weighted avg. follower count of neighbors}
\State \textbf{return} $f$\Comment{The follower count is f}
\EndProcedure
\end{algorithmic}
\end{algorithm}

\vspace{-0.5cm}
\subsection{Evaluation of Follower Count Prediction}
To determine the effectiveness of our method for determining the follower count of a user, we collect a dataset of high quality topic experts on Twitter. Since the customers' follower count is manipulated, we can not establish the ground truth of the correct number of followers for that dataset. However, a genuine user would not indulge in synthetic manipulation of its followers, and hence we choose such users which are topic experts by taking a list of 25 different topics. From this list, we remove the verified users because such users tend to attract a huge number of followers which can also be fake. This might not be a manipulation done by the verified user itself, but result of the natural phenomena of \emph{rich gets richer}. Hence, we collect 500 high quality topic experts for the evaluation of our follower count prediction. Ideally, the displayed follower count on the profile of these users should match our predicted follower count with low error rate. With this assumption, we perform the evaluation of our follower count prediction.

\begin{figure}[h]
\centering
\begin{subfigure}{.20\textwidth}
  \centering
  \includegraphics[width=.9\textwidth]{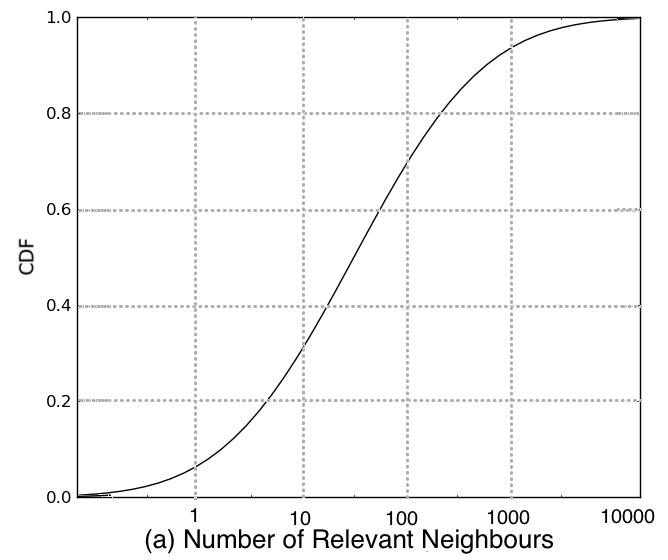}
  \label{fig:relneigh}
\end{subfigure}%
\begin{subfigure}{.28\textwidth}
  \centering
  \includegraphics[width=.9\textwidth]{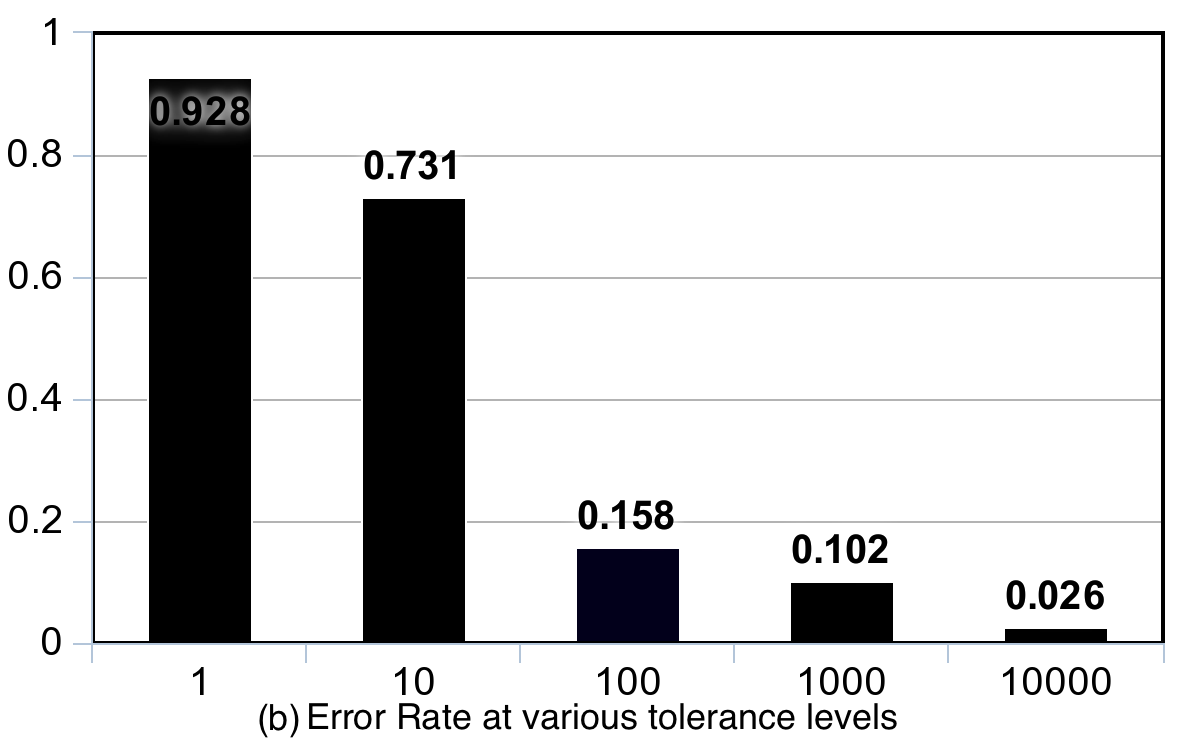}
  \label{fig:evaltol}
\end{subfigure}
\caption{Evaluation of our follower count prediction for expert users with unmanipulated follower count}
\label{fig:evalcountpred}
\end{figure}

Using the dataset of 500 topical experts, we first find their relevant neighborhood. Figure~\ref{fig:evalcountpred}a shows the CDF of number of number of relevant neighbors found for these users. We notice that 70\% users have neighborhood size of 100 or less. Using these neighbors we next predict the follower count. Since our follower count prediction can not be perfect and an exact match to the displayed follower count, we experiment with various tolerance levels, that is the deviation of predicted follower count from the displayed follower count.  Figure~\ref{fig:evalcountpred}b shows the error rate at deviations between $\pm$10 to $\pm$10000 from the displayed follower count. We notice that the error rate is close to zero when we let the predicted follower count deviate from the displayed follower count by  $\pm$10000. However, this is not an acceptable tolerance level, and hence we reduce the tolerance to upto  $\pm$100 and report a maximum error rate of 0.158 ( 84.2\% accuracy). This shows that our follower count prediction method is highly effective.


\vspace{-0.55cm}
\section{Follower Count Estimation and Customer Detection} \label{sec:estimation}
We apply the follower count prediction methodology over a set of customers and find out the real follower count of these users. We use the deviation of estimated follower count from the displayed follower count of users to detect users with manipulated follower count. We are able to find out 98.62\% customers successfully with a low threshold of 10\% deviation between predicted and displayed follower count. 

Since we do not have the ground truth of the real number of followers of these accounts before they indulged in black-market services, we choose a different evaluation metric to assess the effectiveness of our follower prediction methodology. We determine the effectiveness of the follower count prediction methodology by checking its tolerance against manipulation from black-market services. 


\begin{table}[]
\centering
\small
\begin{tabular}{@{}lllllll@{}}
\toprule
   & \multicolumn{2}{l}{Random} & \multicolumn{2}{l}{Customers} & \multicolumn{2}{l}{Controlled} \\ \midrule
   & Klout    & FCP    & Klout     & FCP     & Klout      & FCP     \\ \midrule
G1 & --       & 0.13            & 0.45      & 0.19              & 0.32       & 0.06              \\
G2 & --       & 0.11            & 0.48      & 0.12              & 0.11       & 0.02              \\
G3 & --       & 0.02            & 0.39      & 0.10              & 0.12       & 0.03              \\ \bottomrule
\end{tabular}
\caption{Effect of manipulation on the follower count estimate and Klout score over a period of time. (FCP = using our follower count prediction method). Computed values lies between 0.0 (perfect tolerance) and 1.0 (zero tolerance). The three groups are users divided into buckets with various follower counts. G1$\leq$1000, 1000$<$G2$<$10000, G3$\geq$10000}
\label{tab:evaluation}
\end{table}

For this analysis, we use three datasets. We use the previously used customer dataset D$_{cust}$ and set of random users D$_{rand}$. In addition, we also create 50 dummy Twitter accounts and make these take services from freemium black-markets in our list. We call this dataset D$_{dummy}$. We further divide each dataset into three groups based on its number of followers on Twitter -- G1 with less than 1000 followers, G2 between 1000 and 10,000 followers and G3 with over 10,000 followers. We also compare the tolerance of our follower prediction methodology with a well known influence metric - Klout. Ideally, any influence metric, or our predicted follower count should not change if a user is synthetically manipulating its follower count. Therefore,  we define our evaluation metric as the difference between predicted change (in Klout score, or follower count) and the expected change, divided by the predicted change. This means, that our evaluation metric would show a value of 0.0 in case of perfect tolerance (or perfect prediction of follower count), and a value of 1.0 in case of zero tolerance. Table~\ref{tab:evaluation} shows the effect of manipulation on Klout score and our predicted follower count. For random data, we assume that there is no manipulation over time. Since we do not know the ideal scenario for random users, we do not report the effect on Klout score. We find that our predicted follower count for customers does not change much (remains close to 0.10), though there is a significant jump in the Klout score of the corresponding customers (average 0.47). We see that our follower count prediction shows even higher tolerance in case of dummy accounts with only a small change in the predicted follower count (average 0.011), as compared to high increase in the Klout score of the dummy accounts (0.32 for G1). This shows that unlike other influence metrics, our method to predict follower count is accurate, and also tolerant against manipulation by black-market services.

\vspace{-0.15cm}
\section{Conclusion}
In this study, we take a step forward towards countering manipulation of follower count by precisely predicting the follower count (84.2\% accuracy) of Twitter users, and show that our prediction is tolerant against manipulation to black-markets. The strength of our method lies in the temporal signatures we use for neighborhood detection of a user. Since temporal signatures are hard to manipulate, neighborhood detection using such features ensures capturing similar users in the neighborhood. The predicted follower count of a user is then based on it's neighborhood drawn out from a large sample of random Twitter dataset. We conclude our study by effectively detecting customers (98.62\% precision) using our method, and showing the tolerance of our predicted follower count against manipulation. 


\bibliographystyle{ACM-Reference-Format}
\scriptsize{
\bibliography{sigproc}}


\end{document}